\documentclass[conference]{IEEEtran}
\usepackage{cite}
\usepackage{amsmath,amssymb,amsfonts}
\usepackage{algorithmic}
\usepackage{graphicx}
\usepackage{textcomp}
\usepackage{xcolor}
\usepackage{orcidlink}
\usepackage{glossaries}
\usepackage{listings}
\usepackage{stfloats} 
\usepackage{subcaption} 
\usepackage[outputdir=build,frozencache=true]{minted}
\definecolor{bg}{rgb}{0.95,0.95,0.95}
\usepackage{lipsum}

\usepackage[font=footnotesize]{caption}

\usepackage[T1]{fontenc}
\usepackage{lmodern}

\newacronym[plural=WANs, firstplural={Wide Area Networks (WANs)}]{wan}{WAN}{Wide Area Network}
\newacronym[plural=WSNs, firstplural={Wireless Sensor Networks (WSNs)}]{wsn}{WSN}{Wireless Sensor Network}
\newacronym{simd}{SIMD}{Single Instruction Multiple Data}
\newacronym{os}{OS}{Operating System}
\newacronym{ble}{BLE}{Bluetooth Low-Energy}
\newacronym{wifi}{Wi-FI}{Wireless Fidelity}
\newacronym[plural=DVS, firstplural={Dynamic Vision Sensors (DVS)}]{dvs}{DVS}{Dynamic Vision Sensor}
\newacronym{ptz}{PTZ}{Pan-Tilt Unit}

\newacronym[plural=FLLs,firstplural=Frequency Locked Loops (FLLs)]{fll}{FLL}{Frequency Locked Loop}
\newacronym{dram}{DRAM}{Dynamic Random Access Memory}
\newacronym{fpu}{FPU}{Floating Point Unit}
\newacronym{dma}{DMA}{Direct Memory Access}
\newacronym[plural=LUTs, firstplural={Lookup Tables (LUTs)}]{lut}{LUT}{Lookup Table}
\newacronym[plural=FPGAs, firstplural={Field Programmable Gate Arrays (FPGAs)}]{fpga}{FPGA}{Field Programmable Gate Array}
\newacronym{dsp}{DSP}{Digital Signal Processing}
\newacronym{mcu}{MCU}{Microcontroller Unit}
\newacronym{spi}{SPI}{Serial Peripheral Interface}
\newacronym{cpi}{CPI}{Camera Parallel Interface}
\newacronym{fifo}{FIFO}{First-In First-Out Queue}
\newacronym{uart}{UART}{Universal Asynchronous Receiver-Transmitter}
\newacronym{raw}{RAW}{Read-After-Write}
\newacronym[plural=ISAs, firstplural={Instruction Set Architectures (ISAs)}]{isa}{ISA}{Instruction Set Architecture}
\newacronym{xbar}{XBAR}{crossbar}
\newacronym[firstplural=Scratch-Pad Memories (SPMs)]{spm}{SPM}{Scratch-Pad Memory}
\newacronym{ppa}{PPA}{Power Performance Area}
\newacronym{ipi}{IPI}{Inter-Processor Interrupt}
\newacronym[firstplural=Software-Generated Interrupts (SGIs)]{sgi}{SGI}{Software-Generated Interrupt}
\newacronym{pe}{PE}{Processing Element}
\newacronym{tcdm}{TCDM}{Tightly-Coupled Data Memory}
\newacronym{lsu}{LSU}{Load-Store Unit}
\newacronym{icache}{I\$}{Instruction Cache}
\newacronym{dcache}{D\$}{Data Cache}
\newacronym{wfi}{WFI}{Wait For Interrupt}
\newacronym{gpc}{GPC}{GPU Processing Cluster}
\newacronym{cpu}{CPU}{Central Processing Unit}
\newacronym{gpu}{GPU}{Graphics Processing Unit}
\newacronym{llc}{LLC}{Last-Level Cache}
\newacronym{sm}{SM}{Streaming Multiprocessor}
\newacronym[firstplural=Networks on Chip (NoCs)]{noc}{NoC}{Network on Chip}

\newacronym{ste}{STE}{Straight-Through-Estimator}

\newacronym[plural=PTUs, firstplural={Pan-Tilt Units}]{ptu}{PTU}{Pan-Tilt Unit}
\newacronym{mdf}{MDF}{Medium-density fibreboard}
\newacronym{cvat}{CVAT}{Computer Vision Annotation Tool}
\newacronym{coco}{COCO}{Common Objects in Context}
\newacronym{soa}{SoA}{State of the Art}
\newacronym{sf}{SF}{Sensor Fusion}

\newacronym{dl}{DL}{Deep Learning}
\newacronym{bn}{BN}{Batch Normalization}
\newacronym{FGSM}{FBK}{Fast Gradient Sign Method}
\newacronym{lr}{LR}{Learning Rate}
\newacronym{sgd}{SGD}{Stochastic Gradient Descent}
\newacronym{gd}{GD}{Gradient Descent}

\newacronym{sta}{STA}{Static Timing Analysis}

\newacronym[plural=GPIOs, firstplural={General Purpose Inupt Outputs (GPIOs)}]{gpio}{GPIO}{General Purpose Input Output}
\newacronym[plural=LDOs, firstplural={Low Dropout Regulators (LDOs)}]{ldo}{LDO}{Low Dropout Regulator}

\newacronym{inq}{INQ}{Incremental Network Quantization}

\newacronym{CV}{CV}{Computer Vision}
\newacronym{EoT}{EoT}{Expectation over Transformation}
\newacronym{RPN}{RPN}{Region Proposal Network}
\newacronym{TV}{TV}{Total Variation}
\newacronym{NPS}{NPS}{Non-Printability Score}
\newacronym{STN}{STN}{Spatial Transformer Network}
\newacronym{MTCNN}{MTCNN}{Multi-Task Convolutional Neural Network}
\newacronym{YOLO}{YOLO}{You Only Look Once}
\newacronym{SSD}{SSD}{Single Shot Detector}
\newacronym{SOTA}{SOTA}{State of the Art}
\newacronym{NMS}{NMS}{Non-Maximum Suppression}
\newacronym{ic}{IC}{Integrated Circuit}
\newacronym{rf}{RF}{Radio Frequency}
\newacronym{tcxo}{TCXO}{Temperature Controlled Crystal Oscillator}
\newacronym{jtag}{JTAG}{Joint Test Action Group industry standard}
\newacronym{swd}{SWD}{Serial Wire Debug}
\newacronym{sdio}{SDIO}{Serial Data Input Output}

\newacronym[plural=PCBs, firstplural={Printed Circuit Boards (PCB)}]{pcb}{PCB}{Printed Circuit Board}
\newacronym[plural=ASICs, firstplural={Application Specific Integrated Circuits}]{asic}{ASIC}{Application Specific Integrated Circuit}

\newacronym[plural=BNNs, firstplural={Binary Neural Networks (BNNs)}]{bnn}{BNN}{Binary Neural Network}
\newacronym[plural=NNs, firstplural={Neural Networks}]{nn}{NN}{Neural Network (NNs)}
\newacronym[plural=SCMs, firstplural={Standard Cell Memories (SCMs)}]{scm}{SCM}{Standard Cell Memory}
\newacronym{ann}{ANN}{Artificial Neural Networks}
\newacronym{ml}{ML}{Machine Learning}
\newacronym{ai}{AI}{Artificial Intelligence}
\newacronym{iot}{IoT}{Internet of Things}
\newacronym{fft}{FFT}{Fast Fourier Transform}
\newacronym[plural=OCUs, firstplural={Output Channel Compute Units (OCUs)}]{ocu}{OCU}{Output Channel Compute Unit}
\newacronym{alu}{ALU}{Arithmetic Logic Unit}
\newacronym{mac}{MAC}{Multiply-Accumulate}
\newacronym[firstplural={systems-on-chip (SoCs)}]{soc}{SoC}{system-on-chip}
\newacronym[firstplural={multi-processor systems-on-chip (MPSoCs)}]{mpsoc}{MPSoC}{multi-processor system-on-chip}

\newacronym{PGD}{PGD}{Projected Gradient Descend}
\newacronym{CW}{CW}{Carlini-Wagner}
\newacronym{OD}{OD}{Object Detection}

\newacronym{rrf}{RRF}{RADAR Repetition Frequency}
\newacronym{nlp}{NLP}{Natural Language Processing}
\newacronym{qam}{QAM}{Quadrature Amplitude Modulation}
\newacronym{rri}{RRI}{RADAR Repetition Interval}
\newacronym{radar}{RADAR}{Radio Detection and Ranging}
\newacronym{loocv}{LOOCV}{Leave-one-out cross validation}

\newacronym{bsp}{BSP}{Board Support Package}
\newacronym{ttn}{TTN}{The Things Network}
\newacronym{wip}{WIP}{Work in Progress}
\newacronym{json}{JSON}{JavaScript Object Notation}
\newacronym{qat}{QAT}{Quantization-Aware Training}

\newacronym{cls}{CLS}{Classification Error}
\newacronym{loc}{LOC}{Localization Error}
\newacronym{bkgd}{BKGD}{Background Error}
\newacronym{roc}{ROC}{Receiver Operating Characteristic}
\newacronym{frr}{FRR}{False Rejection Rate}
\newacronym{eer}{EER}{Equal Error Rate}
\newacronym{snr}{SNR}{Signal-to-Noise Ratio}
\newacronym{flop}{FLOP}{Floating-Point Operation}
\newacronym{fp}{FP}{Floating-Point}
\newacronym{fps}{FPS}{Frames Per Second}
\newacronym{oi}{OI}{Operational Intensity}

\newacronym{gsc}{GSC}{Google Speech Commands}
\newacronym{mswc}{MSWC}{Multilingual Spoken Words Corpus}
\newacronym{demand}{DEMAND}{Diverse Environments Multichannel Acoustic Noise Database}

\newacronym[plural=SNNs, firstplural={Spiking Neural Networks (SNNs)}]{snn}{SNN}{Spiking Neural Network}
\newacronym[plural=DNNs, firstplural={Deep Neural Networks (DNNs)}]{dnn}{DNN}{Deep Neural Network}
\newacronym[plural=TCNs,firstplural=Temporal Convolutional Networks]{tcn}{TCN}{Temporal Convolutional Network}
\newacronym[plural=CNNs,firstplural=Convolutional Neural Networks (CNNs)]{cnn}{CNN}{Convolutional Neural Network}
\newacronym[plural=TNNs,firstplural=Ternarized Neural Networks]{tnn}{TNN}{Ternarized Neural Network}
\newacronym{ds-cnn}{DS-CNN}{Depthwise Separable Convolutional Neural Network}
\newacronym{rnn}{RNN}{Recurrent Neural Network}
\newacronym{gcn}{GCN}{Graph Convolutional Network}
\newacronym{mhsa}{MHSA}{Multi-Head Self Attention}
\newacronym{crnn}{CRNN}{Convolutional Recurrent Neural Network}
\newacronym{clca}{CLCA}{Convolutional Linear Cross-Attention}

\newacronym{bf}{BF}{Beamforming}
\newacronym{anc}{ANC}{Active Noise Cancellation}
\newacronym{agc}{AGC}{Automatic Gain Control}
\newacronym{se}{SE}{Speech Enhancement}
\newacronym{mct}{MCT}{Multi-Condition Training}
\newacronym{mcta}{MCTA}{Multi-Condition Training \& Adaptation}
\newacronym{pcen}{PCEN}{Per-Channel Energy Normalization}
\newacronym{mfcc}{MFCC}{Mel-Frequency Cepstral Coefficient}
\newacronym{asr}{ASR}{Automated Speech Recognition}
\newacronym{kws}{KWS}{Keyword Spotting}
\newacronym{odl}{ODL}{On-Device Learning}

\newacronym{nl-kws}{NL-KWS}{Noiseless Keyword Spotting}
\newacronym{na-kws}{NA-KWS}{Noise-Aware Keyword Spotting}
\newacronym{odda}{ODDA}{On-Device Domain Adaptation}
\newacronym{hpm}{HPM}{High-Performance Mode}
\newacronym{lpm}{LPM}{Low-Power Mode}

\newcommand{\ResultEightByEightCrossbarOverheadkGE}{13.1}
\newcommand{\ResultEightByEightCrossbarOverheadPercent}{9}
\newcommand{\ResultSixteenBySixteenCrossbarOverheadkGE}{45.4}
\newcommand{\ResultSixteenBySixteenCrossbarOverheadPercent}{12}

\newcommand{\ResultSixteenBySixteenCrossbarFrequencyOverheadPercent}{6}
\newcommand{\ResultThirtyTwoClusterEightKiBParallelFraction}{97}
\newcommand{\ResultThirtyTwoClusterTwoKiBSpeedup}{13.5}
\newcommand{\ResultThirtyTwoClusterThirtyTwoKiBSpeedup}{16.2}
\newcommand{\ResultThirtyTwoClusterGeometricMeanSpeedup}{5.6}
\newcommand{\ResultBaselineTileNOperationalIntensity}{1.9}
\newcommand{\ResultBaselineTileNPerformanceGFLOPS}{114.4}
\newcommand{\ResultBaselineTileNPerformancePercentage}{92}
\newcommand{\ResultHybridTileNOperationalIntensityIncrease}{3.7}
\newcommand{\ResultHybridTileNPerformanceIncrease}{2.6}
\newcommand{\ResultMulticastTileNOperationalIntensityIncrease}{16.5}
\newcommand{\ResultMulticastTileNPerformanceIncrease}{3.4}
\newcommand{\ResultMulticastTileNPerformanceIncreaseOverHybridPercentage}{29}
\newcommand{\ResultMulticastTileNPerformanceGFLOPS}{391.4}

\setminted[]{
    fontsize=\footnotesize,
    bgcolor=bg,
    style=tango,
    breaklines
}

\def\BibTeX{{\rm B\kern-.05em{\sc i\kern-.025em b}\kern-.08em
    T\kern-.1667em\lower.7ex\hbox{E}\kern-.125emX}}
\begin{document}

\title{A Multicast-Capable AXI Crossbar\\ for Many-core Machine Learning Accelerators}

\ifdefined\blindreview
\else
\author{\IEEEauthorblockN{Luca Colagrande}
\IEEEauthorblockA{\textit{Integrated
Systems Laboratory (IIS)} \\
\textit{ETH Zurich}\\
Zurich, Switzerland \\
colluca@iis.ee.ethz.ch
\orcidlink{0000-0002-7986-1975}}
\and
\IEEEauthorblockN{Luca Benini}
\IEEEauthorblockA{\textit{Integrated
Systems Laboratory (IIS)} \\
\textit{ETH Zurich}\\
Zurich, Switzerland \\
lbenini@iis.ee.ethz.ch
\orcidlink{0000-0001-8068-3806}}
}
\fi

\maketitle

\begin{abstract}
To keep up with the growing computational requirements of machine learning workloads, many-core accelerators integrate an ever-increasing number of processing elements, putting the efficiency of memory and interconnect subsystems to the test. In this work, we present the design of a multicast-capable AXI crossbar, with the goal of enhancing data movement efficiency in massively parallel machine learning accelerators. We propose a lightweight, yet flexible, multicast implementation, with a modest area and timing overhead (12\,\% and 6\,\% respectively) even on the largest physically-implementable 16-to-16 AXI crossbar. To demonstrate the flexibility and end-to-end benefits of our design, we integrate our extension into an open-source 288-core accelerator. We report tangible performance improvements on a key computational kernel for machine learning workloads, matrix multiplication, measuring a 29\,\% speedup on our reference system.
\end{abstract}

\begin{IEEEkeywords}
AI accelerators, on-chip networks, multicast communication, AXI
\end{IEEEkeywords}

\section{Introduction}

In recent years, a wide range of many-core general-purpose accelerators have emerged to keep up with the computational requirements of modern \gls{ml} workloads \cite{peng2024}.
Aiming for higher peak performance figures, these accelerators integrate an ever-increasing number of \glspl{pe}:
the number of CUDA cores in Nvidia's leading \glspl{gpu} increased by more than $2\,\times$ in only two years, rising from 6912 in the A100 \cite{choquette2021} to 16896 in the H100 \cite{choquette2023}.

To translate peak performance into actual performance, it is critical to keep all \glspl{pe} busy for a significant fraction of the operating time.
This poses a significant challenge on the memory and interconnect subsystems, which must be able to sustain the bandwidth required to feed the \glspl{pe} with data.
Pressure on main memory can be relieved by reusing data on-chip.
To this end, most accelerators present a \gls{llc}; a notable example is Nvidia's H100 \gls{gpu} with its 50~MB L2 cache \cite{choquette2023}.

To further multiply on-chip bandwidth, most accelerators feature additional levels of memory, e.g. shared memory in \gls{gpu} \glspl{sm}, and on-chip networks to provide shorter and parallel communication paths between \glspl{pe}.
Emblematically, Nvidia also recently introduced direct \gls{sm}-to-\gls{sm} communication within \glspl{gpc} in their Hopper-architecture \glspl{gpu} \cite{choquette2023}, where \glspl{sm} in a \gls{gpc} are interconnected together by a dedicated on-chip \gls{sm}-to-\gls{sm} network.

These parallel communication paths can be exploited by taking advantage of a computation's data reuse patterns.
In the case of matrix multiplication $\textbf{C}=\textbf{A}\times\textbf{B}$, a key kernel for \gls{ml} workloads, blocks of rows of matrix \textbf{A} are loaded into distinct clusters, while blocks of columns of matrix \textbf{B} have to be broadcast to all clusters (as detailed in section \ref{sec:performance}).
In this setting, multicast communication is extremely beneficial; as such, many recent commercial platforms \cite{prabhakar2024, vasiljevic2024, maddury2024, makino2024} integrate multicast-capable on-chip networks, but their implementations remain undisclosed and, to the best of our knowledge, there are no detailed performance analyses of these designs in the open literature.

Most works in the literature focus on the design of multicast-capable networks for cache-coherent shared-memory systems \cite{jerger2008, abad2009, krishna2011, konstantinou2020}, employing multicast in the coherency protocol implementation. For area and energy efficiency reasons, massively parallel \gls{ml} accelerators do not typically implement cache-coherency, relying on software-managed \glspl{spm} instead; \glspl{gpu} being a prominent example with their \glspl{sm}' shared memories.
Other works either assume a mesh topology \cite{krishna2011, ouyang2021, ouyang2023}, implement destination encodings which are not scalable to the massive parallelism in \gls{ml} accelerators \cite{abad2009, kim2010, zuckerman2024}, or both \cite{wang2009, samman2008, ma2012}.

This work presents the design of a scalable multicast-capable \gls{xbar}, suited for the implementation of on-chip networks for massively parallel \gls{ml} accelerators.
It further differentiates from previous works in that it is fully AXI-compliant and open-source, and thus readily available for integration with standard IPs.
Finally, to the best of our knowledge, this is the first work to evaluate the benefits of multicast communication on a key \gls{ml} kernel. To summarize our contributions, we:
\begin{enumerate}
    \item Design and implement a multicast-capable AXI \gls{xbar}, releasing it as open-source hardware.
\ifdefined\blindreview
    \footnote{www.hidden-for-double-blind-review.com}
\else
    \footnote{https://github.com/colluca/axi/tree/multicast}
\fi
    \item Extend Occamy \cite{occamy}, an existing open-source many-core \gls{ml} accelerator, with multicast capabilities.
\ifdefined\blindreview
    \footnote{www.hidden-for-double-blind-review.com}
\else
    \footnote{https://github.com/colluca/occamy/tree/multicast}
\fi
    \item Evaluate the \gls{xbar}'s area and timing characteristics and the overhead to support multicast.
    \item Evaluate the benefits of multicast communication on a key computational kernel for \gls{ml} workloads.
\end{enumerate}

\noindent
We elaborate on the first two contributions in section \ref{sec:implementation}.
The latter two are discussed in section \ref{sec:results}.

\section{Implementation}
\label{sec:implementation}

\subsection{Multicast-Capable AXI Crossbar}
\label{sec:xbar}

\begin{figure*}[htb]
  \centering

  \begin{subfigure}{0.355\textwidth}
    \centering
    \includegraphics[width=\textwidth]{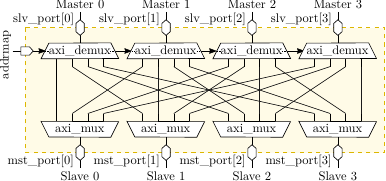}
    \caption{}
    \label{fig:xbar}
  \end{subfigure}
  \hfill
  \begin{subfigure}{0.34\textwidth}
    \centering
    \includegraphics[width=\textwidth]{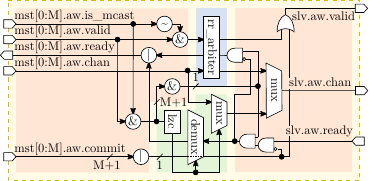}
    \caption{}
    \label{fig:mux}
  \end{subfigure}
  \hfill
  \begin{subfigure}{0.26\textwidth}
    \centering
    \includegraphics[width=\textwidth]{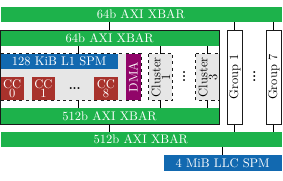}
    \caption{}
    \label{fig:occamy}
  \end{subfigure}

  \vspace{0.4em}

  \begin{subfigure}{0.84\textwidth}
    \centering
    \includegraphics[width=\textwidth]{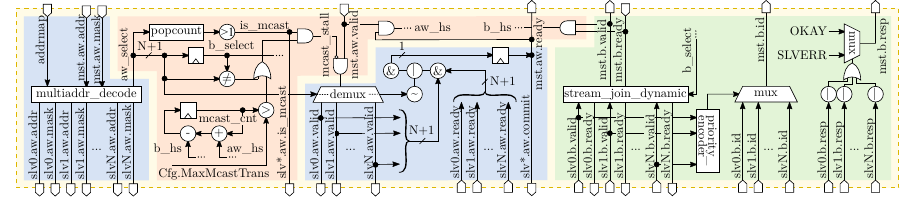}
    \caption{}
    \label{fig:demux}
  \end{subfigure}
  \hfill
  \begin{subfigure}{0.15\textwidth}
    \centering
    \includegraphics[width=\textwidth]{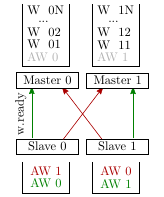}
    \caption{}
    \label{fig:deadlock}
  \end{subfigure}

  \caption{(a) Block diagram of a 4-to-4 AXI \gls{xbar}, (b) AXI mux submodule (unicast datapath is highlighted in blue, multicast datapath in green, and the logic arbitrating the two in orange), (c) Occamy SoC and (d) AXI demux submodule (multicast stall logic is highlighted in orange, logic controlling AW channel forking in blue, and B channel joining in green); (e) Scenario creating the deadlock condition.}
\end{figure*}

We develop our contributions on the open-source AXI \gls{xbar} design by Kurth et al. \cite{kurth2022}.
Its architecture is shown in figure \ref{fig:xbar}.
Masters and slaves are connected through an array of demuxes and muxes.
The \gls{xbar} is associated with an address map: a set of address rules, each mapping an address interval to a slave of the \gls{xbar}.
When a master sends a write request, the address is compared with every rule in the address map (by the address decoder), and the request is routed to the slave associated with the matching rule.
The destination address is propagated, unmodified, in the output request.
As multicast only involves write transactions (AW, W and B channels), we ignore AXI's AR and R channels in the following discussion \cite{axi}.

To define a multicast transaction, a write request must carry multiple destination addresses.
Various multi-address encodings have been proposed in the networking field \cite{chiang1994}, to address multiple nodes in a network.
While our method presents some similarities with the ``multiple region mask'' encoding \cite{chiang1994}, we target the representation of multiple addresses in the global memory space of a system, rather than subcubes of a k-ary n-cube network.

We extend the AXI protocol, without compromising backward compatibility, by passing a mask in the \lstinline{aw_user} signal.
If a bit in the mask is set to 1, the corresponding bit in the address is interpreted as a don't care (X), encoding both logic 0 and 1.
By masking $n$ bits in the address we can represent $2^n$ addresses.
For direct correspondence between address and mask bits, we take the mask to be as wide as the address, although this is not mandatory.

Figure \ref{fig:multiaddr} presents two example address sets that can be represented with our encoding.
While not all possible address sets can be represented, our encoding is suited for massively parallel accelerators, as the encoding size scales logarithmically with the total size of the address space and is independent of the address set size.
Conversely, the ``all destination'' encoding \cite{chiang1994}, which can represent any address set, scales linearly with the address set size.

We extend the address decoder to support multi-address encodings.
The output is a mask (\lstinline{aw_select}) indicating which slaves contain at least one of the destination addresses, together with the subset falling within each slave.

We require every multicast-targetable region, defined by a ``multicast rule'', to 1) be a power-of-two in size and 2) be aligned to an integer multiple of its size.
Any rule satisfying these constraints can be converted from the interval-form encoding (IFE) to the mask-form encoding (MFE), using the following formulas:

\begin{minted}{python}
  mfe.addr = ife.start_addr
  mfe.mask = ife.end_addr - ife.start_addr - 1
\end{minted}

We integrate logic to convert all multicast rules to mask form.
Calculating \lstinline{aw_select} then boils down to:
\begin{minted}{python}
  masked_bits         = req.mask | rule.mask
  match_bits          = ~(req.addr ^ rule.addr)
  aw_select[rule.idx] = &(masked_bits | match_bits)
\end{minted}
The intersection between the request's and a rule's address sets can be found by resolving the masked bits as:
\begin{minted}{python}
  out.mask = req.mask & rule.mask
  out.addr = (~req.mask & req.addr) | (req.mask & rule.addr)
\end{minted}

\begin{figure}[t]
  \captionsetup{belowskip=-1em}
  \centering
  \includegraphics[width=\columnwidth]{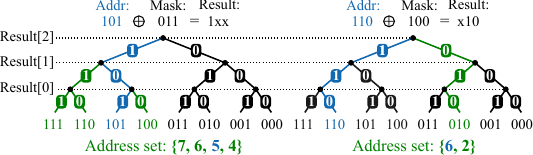}
  \caption{Examples of contiguous (left) and strided (right) address sets representable with our encoding, as paths in the binary number tree. Mask bits selectively fork the path of the original address (blue).}
  \label{fig:multiaddr}
\end{figure}

The \gls{xbar} logic is implemented in the \lstinline{axi_demux} and \lstinline{axi_mux} submodules.
The prior demultiplexes AW and W channel transactions from a master to the addressed slaves, and multiplexes B channel transactions in the opposite direction.
As B responses from different slaves can arrive out-of-order, the demux blocks AW transactions with the same AXI ID as any outstanding transaction, unless directed to the same slave.
To evaluate this condition, it maintains a table of slaves occupied by outstanding transactions, indexed by AXI ID.

Upon a multicast, multiple B responses are expected from different slaves.
Processing multicast transactions out-of-order would require expensive buffering and deadlock-avoidance logic.
We thus disallow multicast transactions until all outstanding unicast transactions have completed and vice versa.
Multiple outstanding multicast transactions are allowed if directed to the same master ports, within a configurable maximum number.

Figure \ref{fig:demux} shows a high-level block diagram of the multicast logic in the \lstinline{axi_demux} submodule.
The green region highlights the logic responsible for joining B responses from different master ports.
The \lstinline{stream_join_dynamic} module ensures that a B handshake is propagated only after receiving a response from every slave.
All responses carry the same ID; we arbitrarily propagate the ID from the first addressed slave using a priority encoder.
On the other hand, the \lstinline{resp} fields may differ and must be properly joined.
As the AXI specification does not cover this scenario, we choose to return a \lstinline{SLVERR} response if any of the responses are either \lstinline{SLVERR} or \lstinline{DECERR}.
We further disallow exclusive multicast transactions, excluding \lstinline{EXOKAY} responses, so the logic boils down to a simple OR-reduction.

Figure \ref{fig:mux} shows a block diagram of the \lstinline{axi_mux} submodule.
Highlighted in green is the logic required to handle multicast transactions.
Two additional 1-bit signals are generated in every demux and routed to every mux in the \gls{xbar}: \lstinline{aw.is_mcast} and \lstinline{aw.commit}.
The prior is used to select between unicast and multicast datapaths; multicast transactions are prioritized, as they have stricter ordering requirements.
The latter is required to prevent deadlocks.

Consider the scenario represented in figure \ref{fig:deadlock}.
Slave 0 receives the AW0 transaction before AW1. According to the AXI specification \cite{axi}, it must thus receive all W0x transactions before any W1x transaction.
On the other hand, slave 1 expects W transactions in the opposite order.
As we cannot buffer all W transactions, we must stall a transaction until all destinations are ready to receive it.
This condition leads to a deadlock, as master 0 waits on \lstinline{w_ready} from slave 1, and master 1 from slave 0.

To prevent this, we force a master to ``acquire'' all slaves at once, breaking Coffman's ``wait for'' condition \cite{coffman1971}.
This is achieved by using a priority-encoder (\lstinline{lzc} module), to ensure consistent master selections across muxes.
When all addressed muxes are ready, the demux asserts the \lstinline{aw.commit} signal, ``releasing'' the muxes in the following cycle.

\subsection{Multicast-Capable \gls{ml} Accelerator}
\label{sec:occamy}

A block diagram of the Occamy \gls{soc} \cite{occamy} is presented in figure \ref{fig:occamy}.
Occamy integrates a configurable number of Snitch clusters \cite{zaruba2021}, each equipped with a 128\,KiB L1 memory and \gls{dma} engine.
Clusters are interconnected through two networks: a narrow 64-bit network for synchronization and control packets issued by the cores' \glspl{lsu}, and a wide 512-bit network shared by the instruction cache and \gls{dma} subsystems.
Both networks are implemented by a two-level hierarchy of \glspl{xbar}.
At the top level, a configurable-size \gls{llc} is connected to the wide network.

Clusters are mapped to consecutive address intervals of size \lstinline{0x40000} starting from address \lstinline{0x01000000}, satisfying the constraints imposed for the definition of multicast targets.

We integrate our extension in every \gls{xbar} of the two networks.
We further extend the Snitch cluster's \gls{lsu} and \gls{dma} engine, to respectively issue multicast interrupts on the narrow network, accelerating synchronization, and data transfers on the wide network, enhancing data movement efficiency, as we will see in section \ref{sec:performance}.

\section{Results}
\label{sec:results}

\subsection{Area and Timing Analysis}

We synthesize the design using Synopsys' Fusion Compiler 2021.06 under
worst-case conditions at 0.72\,V and 125\,\textdegree C in GLOBALFOUNDRIES’ 12LP+ technology, with a 1\,ns clock constraint.

\begin{figure*}[t]
  \centering
  \begin{subfigure}{0.22\textwidth}
    \centering
    \includegraphics[width=\textwidth]{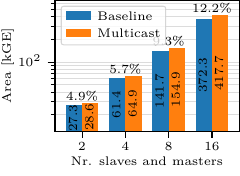}
    \caption{}
    \label{fig:area}
  \end{subfigure}
  \hfill
  \begin{subfigure}{0.28\textwidth}
    \centering
    \includegraphics[width=\textwidth]{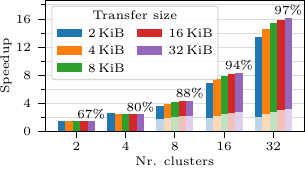}
    \caption{}
    \label{fig:microbenchmark}
  \end{subfigure}
  \hfill
  \begin{subfigure}{0.28\textwidth}
    \centering
    \includegraphics[width=\textwidth]{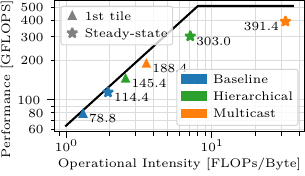}
    \caption{}
    \label{fig:gemm-results}
  \end{subfigure}
  \begin{subfigure}{0.18\textwidth}
    \centering
    \includegraphics[width=\textwidth]{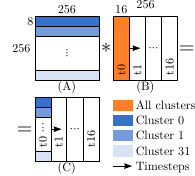}
    \caption{}
    \label{fig:gemm-schedule}
  \end{subfigure}
  \caption{(a) Area of the original and multicast-capable \glspl{xbar} (numbers on top of the bars report the area increase); (b) Speedup on the microbenchmark with our extensions (numbers on top of the bars report the equivalent parallel fraction according to Amdahl's law for the 32\,KiB data points); (c) Performance of the matmul kernel; (d) Parallelization and scheduling of the matmul kernel.}
\end{figure*}

Figure \ref{fig:area} shows the area of an N-to-N \gls{xbar}, with and without multicast support.
On 8-to-8 and 16-to-16 \glspl{xbar}, our extensions introduce overheads of \ResultEightByEightCrossbarOverheadkGE\,kGE and \ResultSixteenBySixteenCrossbarOverheadkGE\,kGE (\ResultEightByEightCrossbarOverheadPercent\,\% and \ResultSixteenBySixteenCrossbarOverheadPercent\,\% of the baseline \gls{xbar}), respectively.
As the area scales quadratically with N, 16-to-16 is typically at the upper limit for \glspl{xbar} that can be implemented at the physical level, and interconnect scale-up is obtained by going multi-stage in a hierarchy of \glspl{xbar} \cite{kurth2022}.

All configurations meet the target 1\,GHz operating frequency, with the exception of the 16-to-16 \gls{xbar} which incurs a very modest \ResultSixteenBySixteenCrossbarFrequencyOverheadPercent\,\% frequency degradation.

\subsection{Performance Evaluation}
\label{sec:performance}

We conduct the performance evaluation through cycle-accurate RTL simulations of the Occamy \gls{soc} using QuestaSim 2023.4, with a 1\,GHz clock frequency. We assume an Occamy system with 32 clusters, organized into 8 groups of 4 clusters each, and a 4\,MiB \gls{llc}.

We first evaluate our extension on a microbenchmark, which consists in one cluster sending the same data to all other clusters using its \gls{dma} engine.
We compare the runtime of the multicast \gls{dma} transfer using our extensions to a multiple-unicast approach, where unicast \gls{dma} transfers are issued to every destination cluster.
For transfers to more than one group (i.e. 8, 16 and 32 clusters), we also compare to a hierarchical software-based multicast approach, where the source cluster sends the data to one cluster in every other group, which in turn forwards the data to the other clusters in its group.
The distribution within groups can thus proceed in parallel.

The colored bars in figure \ref{fig:microbenchmark} show the speedup of the multicast transfer over the multiple-unicast baseline.
The speedup increases with the number of clusters, and approaches the ideal parallel speedup, with the equivalent parallel fraction per Amdahl's law reaching \ResultThirtyTwoClusterEightKiBParallelFraction\,\% on 32 clusters.
This is due to constant sequential overheads, such as the round-trip latency, being amortized over multiple transfers.
Similarly, we observe a small increase in speedup with growing transfer sizes, ranging from \ResultThirtyTwoClusterTwoKiBSpeedup$\times$ to \ResultThirtyTwoClusterThirtyTwoKiBSpeedup$\times$ on a 32-cluster transfer.
The white overlays represent the speedup of the hierarchical software-based multicast approach over the baseline.
As we can see, hardware-supported multicast still gives significant speedups over the software-based approach, with a geometric mean speedup of \ResultThirtyTwoClusterGeometricMeanSpeedup$\times$ on the 32-cluster transfers. 

Finally, we evaluate how multicast support translates to tangible performance improvements on a key computational kernel for \gls{ml} applications, i.e. matrix multiplication (matmul).
We execute the largest square double-precision matrix multiplication tile which fits in Occamy's \gls{llc}: $256\times256$ matrices, accounting for double buffering.
As illustrated in figure \ref{fig:gemm-schedule}, every cluster computes a distinct $8\times256$ row block of the product matrix \textbf{C}, calculating an $8\times16$ tile of the row block at a time.
The corresponding tile of \textbf{A} need only be loaded once into L1, and can be reused in every successive (steady-state) iteration.
The cluster \glspl{dma} are used to move data between \gls{llc} and cluster L1 memories in a double-buffered fashion.

Figure \ref{fig:gemm-results} displays the attained performance in a roofline plot of the Occamy architecture.
The baseline kernel features a low steady-state \gls{oi} of \ResultBaselineTileNOperationalIntensity\,FLOPS/byte, as all clusters have to load the \textbf{B} matrix tile from the \gls{llc}.
This places the kernel in the memory-bound region, achieving \ResultBaselineTileNPerformanceGFLOPS\,GFLOPS, or \ResultBaselineTileNPerformancePercentage\,\% of the maximum theoretical performance with this specific \gls{oi}.

By exploiting multicast, we can load the \textbf{B} matrix tile once and broadcast it to all clusters in parallel. The total number of bytes read from the \gls{llc} is reduced, resulting in \ResultHybridTileNOperationalIntensityIncrease$\times$ and \ResultMulticastTileNOperationalIntensityIncrease$\times$ higher \glspl{oi} respectively with software-based and hardware-supported multicast.
These respectively translate to \ResultHybridTileNPerformanceIncrease$\times$ and \ResultMulticastTileNPerformanceIncrease$\times$ performance improvements, reaching \ResultMulticastTileNPerformanceGFLOPS\,GFLOPS with hardware-supported multicast. This result shows how \gls{ml} applications can benefit from multicast support, making it a viable solution to enhance the on-chip bandwidth utilization of many-core \gls{ml} accelerators.

\section{Conclusion}

In this work, we presented the design of a multicast-capable AXI crossbar, leveraging a scalable, yet flexible, multi-address encoding scheme.
We analyzed the area and timing characteristics of the design, showing how multicast support can be achieved with a modest area and timing overhead (\ResultSixteenBySixteenCrossbarOverheadPercent\,\% and \ResultSixteenBySixteenCrossbarFrequencyOverheadPercent\,\% respectively) even on the largest physically-implementable 16-to-16 AXI crossbar.
We integrated our design into an open-source 288-core \gls{ml} accelerator, demonstrating its flexibility on an actual system.
Finally, we evaluated the performance impact on a key computational kernel for machine learning workloads, matrix multiplication, measuring a \ResultMulticastTileNPerformanceIncreaseOverHybridPercentage\,\% improvement with our solution, proving that multicast can provide a low-cost solution to enhance the on-chip bandwidth utilization of massively parallel \gls{ml} accelerators.



\bibliography{paper}

\begin{thebibliography}{10}
\providecommand{\url}[1]{#1}
\csname url@samestyle\endcsname
\providecommand{\newblock}{\relax}
\providecommand{\bibinfo}[2]{#2}
\providecommand{\BIBentrySTDinterwordspacing}{\spaceskip=0pt\relax}
\providecommand{\BIBentryALTinterwordstretchfactor}{4}
\providecommand{\BIBentryALTinterwordspacing}{\spaceskip=\fontdimen2\font plus
\BIBentryALTinterwordstretchfactor\fontdimen3\font minus
  \fontdimen4\font\relax}
\providecommand{\BIBforeignlanguage}[2]{{%
\expandafter\ifx\csname l@#1\endcsname\relax
\typeout{** WARNING: IEEEtran.bst: No hyphenation pattern has been}%
\typeout{** loaded for the language `#1'. Using the pattern for}%
\typeout{** the default language instead.}%
\else
\language=\csname l@#1\endcsname
\fi
#2}}
\providecommand{\BIBdecl}{\relax}
\BIBdecl

\bibitem{peng2024}
\BIBentryALTinterwordspacing
H.~Peng, C.~Ding, T.~Geng, S.~Choudhury, K.~Barker, and A.~Li, ``Evaluating
  emerging ai/ml accelerators: Ipu, rdu, and nvidia/amd gpus,'' in
  \emph{Companion of the 15th ACM/SPEC International Conference on Performance
  Engineering}, ser. ICPE '24 Companion.\hskip 1em plus 0.5em minus 0.4em\relax
  New York, NY, USA: Association for Computing Machinery, 2024, p. 14–20.
  [Online]. Available: \url{https://doi.org/10.1145/3629527.3651428}
\BIBentrySTDinterwordspacing

\bibitem{choquette2021}
J.~Choquette, W.~Gandhi, O.~Giroux, N.~Stam, and R.~Krashinsky, ``Nvidia a100
  tensor core gpu: Performance and innovation,'' \emph{IEEE Micro}, vol.~41,
  no.~2, pp. 29--35, 2021.

\bibitem{choquette2023}
J.~Choquette, ``Nvidia hopper h100 gpu: Scaling performance,'' \emph{IEEE
  Micro}, vol.~43, no.~3, pp. 9--17, 2023.

\bibitem{prabhakar2024}
\BIBentryALTinterwordspacing
R.~Prabhakar, ``Sambanova sn40l rdu: Breaking the barrier of trillion+
  parameter scale gen ai computing,'' in \emph{2024 IEEE Hot Chips 36 Symposium
  (HCS)}.\hskip 1em plus 0.5em minus 0.4em\relax Los Alamitos, CA, USA: IEEE
  Computer Society, aug 2024, pp. 1--24. [Online]. Available:
  \url{https://doi.ieeecomputersociety.org/10.1109/HCS61935.2024.10664717}
\BIBentrySTDinterwordspacing

\bibitem{vasiljevic2024}
\BIBentryALTinterwordspacing
J.~Vasiljevic and D.~Capalija, ``Blackhole \&amp; tt-metalium: The standalone
  ai computer and its programming model,'' in \emph{2024 IEEE Hot Chips 36
  Symposium (HCS)}.\hskip 1em plus 0.5em minus 0.4em\relax Los Alamitos, CA,
  USA: IEEE Computer Society, aug 2024, pp. 1--30. [Online]. Available:
  \url{https://doi.ieeecomputersociety.org/10.1109/HCS61935.2024.10664810}
\BIBentrySTDinterwordspacing

\bibitem{maddury2024}
\BIBentryALTinterwordspacing
M.~Maddury, P.~Kansal, and O.~Wu, ``Next gen mtia -recommendation inference
  accelerator,'' in \emph{2024 IEEE Hot Chips 36 Symposium (HCS)}.\hskip 1em
  plus 0.5em minus 0.4em\relax Los Alamitos, CA, USA: IEEE Computer Society,
  aug 2024, pp. 1--27. [Online]. Available:
  \url{https://doi.ieeecomputersociety.org/10.1109/HCS61935.2024.10665192}
\BIBentrySTDinterwordspacing

\bibitem{makino2024}
\BIBentryALTinterwordspacing
J.~Makino, ``Mn-core 2: Second-generation processor of mn-core architecture for
  ai and general-purpose hpc application,'' in \emph{2024 IEEE Hot Chips 36
  Symposium (HCS)}.\hskip 1em plus 0.5em minus 0.4em\relax Los Alamitos, CA,
  USA: IEEE Computer Society, aug 2024, pp. 1--22. [Online]. Available:
  \url{https://doi.ieeecomputersociety.org/10.1109/HCS61935.2024.10664802}
\BIBentrySTDinterwordspacing

\bibitem{jerger2008}
N.~E. Jerger, L.-S. Peh, and M.~Lipasti, ``Virtual circuit tree multicasting: A
  case for on-chip hardware multicast support,'' in \emph{2008 International
  Symposium on Computer Architecture}, 2008, pp. 229--240.

\bibitem{abad2009}
P.~Abad, V.~Puente, and J.-A. Gregorio, ``Mrr: Enabling fully adaptive
  multicast routing for cmp interconnection networks,'' in \emph{2009 IEEE 15th
  International Symposium on High Performance Computer Architecture}, 2009, pp.
  355--366.

\bibitem{krishna2011}
T.~Krishna, L.-S. Peh, B.~M. Beckmann, and S.~K. Reinhardt, ``Towards the ideal
  on-chip fabric for 1-to-many and many-to-1 communication,'' in \emph{2011
  44th Annual IEEE/ACM International Symposium on Microarchitecture (MICRO)},
  2011, pp. 71--82.

\bibitem{konstantinou2020}
D.~Konstantinou, C.~Nicopoulos, J.~Lee, G.~C. Sirakoulis, and
  G.~Dimitrakopoulos, ``Smartfork: Partitioned multicast allocation and
  switching in network-on-chip routers,'' in \emph{2020 IEEE International
  Symposium on Circuits and Systems (ISCAS)}, 2020, pp. 1--5.

\bibitem{ouyang2021}
\BIBentryALTinterwordspacing
Y.~Ouyang, F.~Tang, C.~Hu, W.~Zhou, and Q.~Wang, ``Mmnnn: A tree-based
  multicast mechanism for noc-based deep neural network accelerators,''
  \emph{Microprocess. Microsyst.}, vol.~85, no.~C, sep 2021. [Online].
  Available: \url{https://doi.org/10.1016/j.micpro.2021.104242}
\BIBentrySTDinterwordspacing

\bibitem{ouyang2023}
\BIBentryALTinterwordspacing
Y.~Ouyang, J.~Wang, C.~Sun, Q.~Wang, and H.~Liang, ``Urmp: using reconfigurable
  multicast path for noc-based deep neural network accelerators,'' \emph{The
  Journal of Supercomputing}, vol.~79, no.~13, p. 14827–14847, 2023.
  [Online]. Available: \url{https://doi.org/10.1007/s11227-023-05255-7}
\BIBentrySTDinterwordspacing

\bibitem{kim2010}
J.-Y. Kim, J.~Park, S.~Lee, M.~Kim, J.~Oh, and H.-J. Yoo, ``A 118.4 gb/s
  multi-casting network-on-chip with hierarchical star-ring combined topology
  for real-time object recognition,'' \emph{IEEE Journal of Solid-State
  Circuits}, vol.~45, no.~7, pp. 1399--1409, 2010.

\bibitem{zuckerman2024}
\BIBentryALTinterwordspacing
J.~Zuckerman, J.-D. Wellman, A.~Vanamali, M.~Shankar, G.~Tombesi,
  K.~Swaminathan, K.~Lee, M.~Kapur, R.~Philhower, P.~Bose, and L.~P. Carloni,
  ``Towards generalized on-chip communication for programmable accelerators in
  heterogeneous architectures,'' 2024. [Online]. Available:
  \url{https://arxiv.org/abs/2407.04182}
\BIBentrySTDinterwordspacing

\bibitem{wang2009}
L.~Wang, Y.~Jin, H.~Kim, and E.~J. Kim, ``Recursive partitioning multicast: A
  bandwidth-efficient routing for networks-on-chip,'' in \emph{2009 3rd
  ACM/IEEE International Symposium on Networks-on-Chip}, 2009, pp. 64--73.

\bibitem{samman2008}
F.~A. Samman, T.~Hollstein, and M.~Glesner, ``Multicast parallel pipeline
  router architecture for network-on-chip,'' in \emph{2008 Design, Automation
  and Test in Europe}, 2008, pp. 1396--1401.

\bibitem{ma2012}
S.~Ma, N.~E. Jerger, and Z.~Wang, ``Supporting efficient collective
  communication in nocs,'' in \emph{IEEE International Symposium on
  High-Performance Comp Architecture}, 2012, pp. 1--12.

\bibitem{occamy}
G.~Paulin, P.~Scheffler, T.~Benz, M.~Cavalcante, T.~Fischer, M.~Eggimann,
  Y.~Zhang, N.~Wistoff, L.~Bertaccini, L.~Colagrande, G.~Ottavi, F.~K.
  Gürkaynak, D.~Rossi, and L.~Benini, ``Occamy: A 432-core 28.1 dp-gflop/s/w
  83
  stencil and sparse linear algebra computations with 8-to-64-bit
  floating-point support in 12nm finfet,'' in \emph{2024 IEEE Symposium on VLSI
  Technology and Circuits (VLSI Technology and Circuits)}, 2024, pp. 1--2.

\bibitem{kurth2022}
A.~Kurth, W.~Rönninger, T.~Benz, M.~Cavalcante, F.~Schuiki, F.~Zaruba, and
  L.~Benini, ``An open-source platform for high-performance non-coherent
  on-chip communication,'' \emph{IEEE Transactions on Computers}, vol.~71,
  no.~8, pp. 1794--1809, 2022.

\bibitem{axi}
\BIBentryALTinterwordspacing
\emph{AMBA® AXI and ACE Protocol Specification Issue H}, Arm, March 2020.
  [Online]. Available: \url{https://developer.arm.com/documentation/ihi0022/h/}
\BIBentrySTDinterwordspacing

\bibitem{chiang1994}
C.-M. Chiang and L.~M. Ni, ``Multi-address encoding for multicast,'' in
  \emph{Proceedings of the First International Workshop on Parallel Computer
  Routing and Communication}, ser. PCRCW '94.\hskip 1em plus 0.5em minus
  0.4em\relax Berlin, Heidelberg: Springer-Verlag, 1994, p. 146–160.

\bibitem{coffman1971}
\BIBentryALTinterwordspacing
E.~G. Coffman, M.~Elphick, and A.~Shoshani, ``System deadlocks,'' \emph{ACM
  Comput. Surv.}, vol.~3, no.~2, p. 67–78, jun 1971. [Online]. Available:
  \url{https://doi.org/10.1145/356586.356588}
\BIBentrySTDinterwordspacing

\bibitem{zaruba2021}
F.~Zaruba, F.~Schuiki, T.~Hoefler, and L.~Benini, ``Snitch: A tiny pseudo
  dual-issue processor for area and energy efficient execution of
  floating-point intensive workloads,'' \emph{IEEE Transactions on Computers},
  vol.~70, no.~11, pp. 1845--1860, 2021.

\end{thebibliography}
\bibliographystyle{IEEEtran}

\end{document}